\documentclass[fleqn,usenatbib]{mnras}

\usepackage[T1]{fontenc}

\usepackage{graphicx}	
\usepackage{amsmath}	
\usepackage{amssymb}	
\usepackage{rotating}

\usepackage{pdflscape}
\usepackage{soul}

\title[Discovery of new star clusters]{
New star clusters discovered towards the Galactic bulge direction using {\textit{Gaia}} DR2}

\author[F. A. Ferreira et al.]{
F. A. Ferreira$^{1}$\thanks{E-mail: filipe1906@ufmg.br},
W. J. B. Corradi$^{4,1}$,
F. F. S. Maia$^{2}$,
M. S. Angelo$^{3}$  \newauthor
\ and J. F. C. Santos Jr.$^{1}$
\\
$^{1}$Universidade Federal de Minas Gerais, Departamento de F\'isica, Av. Ant\^onio Carlos 6627, 31270-901, Brazil \\
$^{2}$Universidade Federal do Rio de Janeiro, Instituto de F\'isica, 21941-972, Brazil 
\\
$^{3}$Centro Federal de Educa\c c\~ao Tecnol\'ogica de Minas Gerais, Av. Monsenhor Luiz de Gonzaga, 103, 37250-000, Brazil\\
$^{4}$Laborat\'orio Nacional de Astrof{\'{\i}}sica, R. Estados Unidos, 154, 37504-364, Itajub\'a, MG, Brazil
}

\date{Accepted XXX. Received YYY; in original form ZZZ}

\pubyear{2020}

\begin{document}
\label{firstpage}
\pagerange{\pageref{firstpage}--\pageref{lastpage}}
\maketitle

\begin{abstract}

We report the discovery of 34 new open clusters and candidates as a result of a systematic search carried out in 200 adjacent fields of 1\,$\times$\,1\,deg$ ^2$ area projected towards the Galactic bulge, using {\it Gaia} DR2 data. The objects were identified and characterized by a joint analysis of their photometric, kinematic and spatial distribution, which has been consistently used and proved to be effective in our previous works. The discoveries were validated by cross-referencing the objects position and astrometric parameters with the available literature. Besides their coordinates and astrometric parameters, we also provide sizes, ages, distances and reddening for the discovered objects. In particular, 32 clusters are closer than 2 kpc from the Sun, which represents an increment of nearly $39\%$ of objects with astrophysical parameters determined in the nearby inner disk.
Although these objects fill an important gap in the open clusters distribution along the Sagittarius arm, this arm, traced by known clusters, appears to be interrupted, which may be an artifact due to the incompleteness of the cluster census.

\end{abstract}




\section{Introduction}


Finding and characterizing objects projected towards dense fields, like the Galactic bulge, has always been a challenging task. However, in the recent years the \textit{Gaia} DR2 catalogue \citep{Brown:2018}, which provides precise astrometric and photometric data for a billion stars in the whole sky, has allowed a better characterization of open clusters (OCs) \citep[e.g.][]{2019MNRAS.490.1383R,10.1093/mnras/staa3192,2020MNRAS.499.1874M} and revolutionized the search for new objects in the astrometric space, leading to a discovery of hundreds of new OCs in our Galaxy \citep{lp19,cjl20,2020arXiv201014870H}.

Increasing and completing the database of known Galactic OCs with well-determined properties is very important, since clusters have long been used as probes to investigate several Galactic features, such as the disc structure and its scale height \citep[][hereafter CG+2020]{2020A&A...640A...1C}, spiral arms and Galactic rotation curve \citep{2019MNRAS.486.5726D}, metallicity and abundance gradients \citep{2017A&A...603A...2M}. Despite this, it has been shown that the nearby OC census is still incomplete inside 1.8 kpc \citep{cks19}, demanding a continuous search for missing clusters.

  


In previous works, we have found 28 new OCs  by adopting a methodology involving iterative inspection of proper motion and sky charts after applying a series of filters to enhance cluster/field contrast (\citealp[hereafter F+2020]{2019MNRAS.483.5508F,10.1093/mnras/staa1684}). We also characterized these newly discovered clusters, thus helping to expand the parameters database of known OCs in the Galaxy.

In this work, we have applied the same methodology, extending our search to a region of the sky around the Galactic centre hoping to increase the present database of the OC population towards the Galactic bulge.  


This work is organized as follows: in Sect. 2 we describe the method applied to detect and validate the discovery of the new OCs. The analysis is presented in Sect. 3, including the membership assessment and the determination of the astrophysical parameters. In Sect. 4, we compare our discoveries with the known OCs and discuss the main results. The concluding remarks are given in Sect. 5.


\section{Method: Searching for star clusters}
\label{sect:search}

\subsection{Data}
We have used data from the \textit{Gaia} DR2 catalogue \citep{Lindegren:2018,Evans:2018} to search and characterize star clusters towards the Galactic centre direction.  This catalogue provides astrometric (positions, proper motions and parallaxes) and photometric (magnitudes in the $G$, $G_{BP}$ and $G_{RP}$ bands) data for more than 1.3 billion sources \citep{Brown:2018}. Gaia@AIP (https://gaia.aip.de/) online services have been used to extract \textit{Gaia} DR2 data within Galactic coordinates  $-5^{\circ}$ $\leq b \leq$ $5^{\circ}$ and $-10^{\circ}$ $\leq \ell \leq$10$^{\circ}$.


We applied a first basic filter in our data by keeping only stars brighter than $G=18\,$mag, the same filter applied by \cite{cjv18}. This filter removes very faint and less informative sources (with high astrometric uncertainties). A second set of filters, which assures quality to the data by cleaning the sample from contamination due to double stars, astrometric effects from binary stars and calibration problems, was applied by using equations (1), (2) and (3) from \cite{Arenou:2018}. 

\subsection{Tiling and searching regions}

\begin{figure}
\centering
\includegraphics[width=0.98\linewidth]{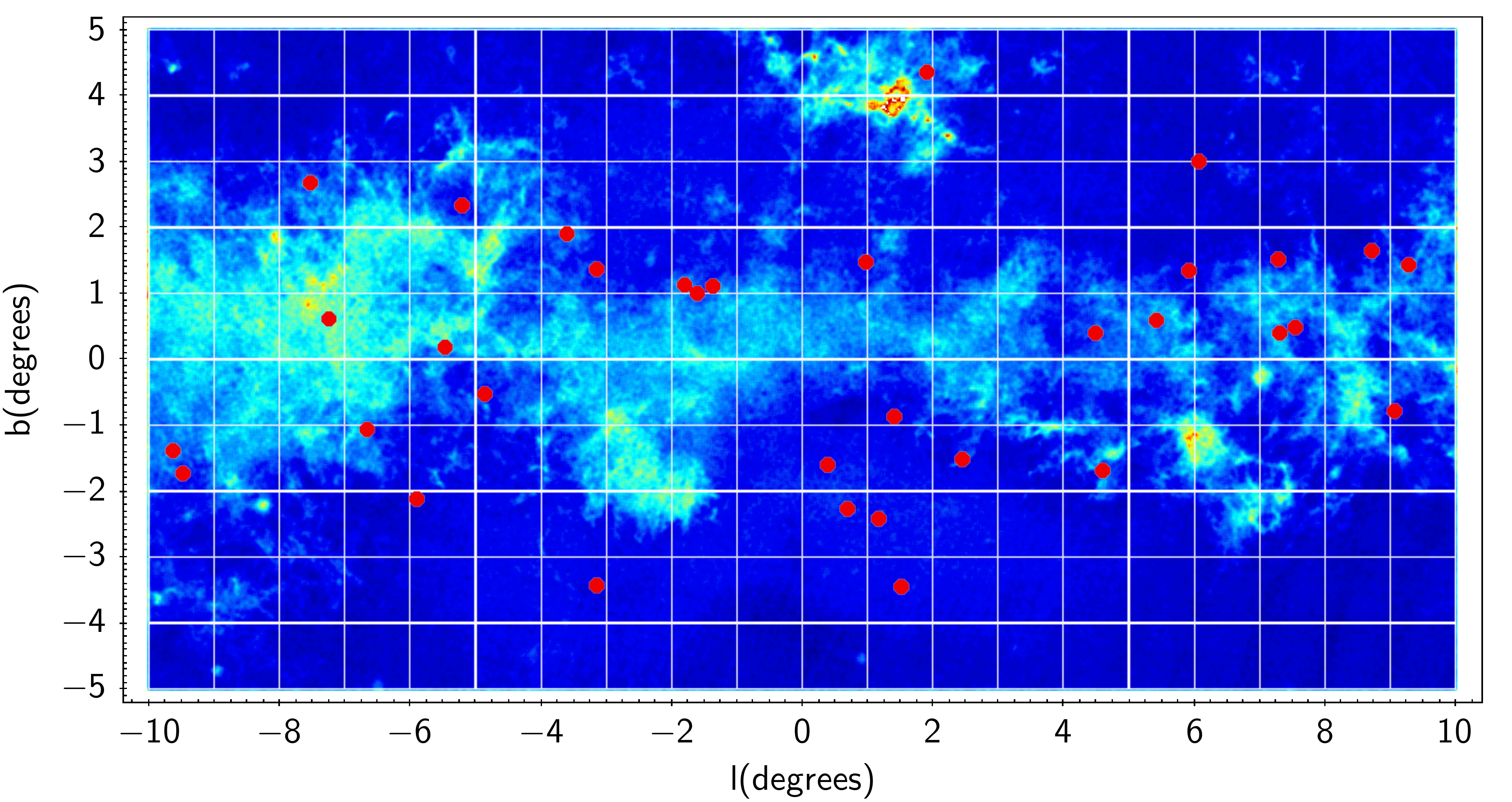}
\caption{Spatial coverage of the Galactic fields surveyed in this work. The size of the white squares indicates the surveyed regions.  The colours indicate relative stellar densities from the {\it Gaia} DR2 catalogue. The red dots indicate the position of the 34 newly discovered OCs in this work.}
\label{fig:tiles_menores}
\end{figure}

In the present work, we have searched for new clusters following procedures similar to those described in F+2020, with the whole surveyed area sectioned in tiles of equal area. We restricted our sample to 200 square tiles of 1\,$\times$\,1\,deg$ ^2$ area in Galactic coordinates (Fig. \ref{fig:tiles_menores}).
 
 
Briefly, for each tile, we determined a mean colour value $G_{BP}-G_{RP}$ and built two smaller subsamples: one with stars bluer than this cut limit and another one with stars redder than this value. 
Both subsamples were analysed on VPDs, where we searched for overdensities by inspecting the stars distribution. Any overdensities found were extracted by applying a box-shaped mask of size 1\,mas\,yr$^{-1}$ 
and searched for clustered structures in sky charts. Any clustered structure was spatially trimmed by applying a circle-shaped mask of 5-10 arcmin radius, depending on its visual size. 
Finally, we inspected the colour-magnitude diagram (CMD) and parallax distribution of the resulting samples to confirm the candidate.

The subsamples filtered by proper motion and spatial masks were used to compute the initial mean values of $\mu_{\alpha}^{*}$, $\mu_{\delta}$, $\varpi$, $\ell$, and $b$ and their dispersion (except for $\ell$ and $b$). These quantities are interactively refined during the analysis of confirmed candidates; e.g. the centre and radius are updated after construction of the radial density profiles, proper motion and parallaxes are updated after assessing membership, etc.



   
In total, we have surveyed a projected area of $\sim$200 square degrees, finding in the first moment a total of 131 objects, which includes known OCs and globular clusters (GCs), repeated signature of the same object in two adjacent tiles and substructure of other objects that were computed as a candidate. To exclude repeated objetcs, we carefully performed an internal match  with the initial catalogue, so this initial number has been reduced to 118 objects (109 OCs and 9 GCs).    

\subsection{Identifying OCs candidates}

To identify the detected OCs, we initially matched our centre coordinates with those on the literature. Then we built a reference database by compiling the following catalogues: \cite{Roser:2016}, \cite{cjl18},  \cite{2018MNRAS.481.3902B}, \cite{2018RAA....18...32G}, \cite{cjl19}, \cite{2019AJ....157...12B} , \cite{2019MNRAS.483.5508F}, \cite{cks19},  \cite{2019MNRAS.484.2181T},  \cite{refId0}, \cite{sla19},  \cite{lp19},\cite{cjl20}, \cite{2020PASP..132c4502H}, F+2020, \cite{2020arXiv200807164Q} and   \cite{2020arXiv201014870H}. We searched for close neighbours within 1$^{\circ}$
and considered a newfound candidate when none of the objects share the same astrometric space or when the centres distance are larger than 5 arcmin for smaller objects and 10 arcmin for larger objects. At this point, we identified 84 known objects (including OCs and bulge GCs). 



After this first procedure, we determined the limiting radius for the candidates and repeated the matching procedure, but now taking into account the sizes of the objects. So, if the separation of the centres is larger than or equal the sum of our limiting radius and the literature radius, we assume it as a new cluster, because we do not expect any contamination. Finally, we also cross-matched our clusters member lists with the members available in the literature. If the cluster and its neighbours have an area in common  in the sky -- for example a halo overlap --, but do not present any shared members, we assume they are different objects. For the best of our knowledge, the 34 new clusters reported in this work do not share stars nor are sub-structures of any known objects.

\begin{table*}
\caption{Derived astrophysical parameters for the studied clusters. Uncertainties are also presented, except in the case of $\mu_{\alpha}^{*}$, $\mu_{\delta}$ and $\varpi$, where 1-sigma dispersions are reported instead.}
\scriptsize
\label{Tab:clusters_prop1}
\begin{tabular}{l r r r@{$\,\pm\,$}l r@{$\,\pm\,$}l r@{$\,\pm\,$}l  r@{$\,\pm\,$}l r@{$\,\pm\,$}l r@{$\,\pm\,$}l r@{$\,\pm\,$}l}
\hline
\multicolumn{1}{c}{ID} &
\multicolumn{1}{c}{RA} &
\multicolumn{1}{c}{DEC} &
\multicolumn{2}{c}{$r_\mathrm{lim}$} &
\multicolumn{2}{c}{E(B$-$V)} &
\multicolumn{2}{c}{$\mathrm{(m-M)}_0$} &
\multicolumn{2}{c}{$\log{t}$} &
\multicolumn{2}{c}{$\mu_{\alpha}^{*}$} &
\multicolumn{2}{c}{$\mu_{\delta}$} &
\multicolumn{2}{c}{$\varpi$} \\

\multicolumn{1}{c}{} &
\multicolumn{1}{c}{$(deg)$} &
\multicolumn{1}{c}{$(deg)$} &
\multicolumn{2}{c}{$(arcsec)$} &
\multicolumn{2}{c}{$(mag)$} &
\multicolumn{2}{c}{$(mag)$} &
\multicolumn{2}{c}{} &
\multicolumn{2}{c}{$(mas/yr)$} &
\multicolumn{2}{c}{$(mas/yr)$} &
\multicolumn{2}{c}{$(mas)$} \\
\hline

  UFMG63 & 263.400 & -25.010 & 654.2 & 19.1 & 1.40 & 0.05 & 10.00 & 0.15 & 7.80 & 0.20 & 1.231 & 0.102 & -0.279 & 0.097 & 0.805 & 0.062\\
  UFMG64 & 267.055 & -22.186 & 816.7 & 62.9 & 0.65 & 0.08 & 10.55 & 0.20 & 8.15 & 0.25 & 2.225 & 0.149 & -0.329 & 0.101 & 0.629 & 0.064\\
  UFMG65 & 258.947 & -33.750 & 441.7 & 62.9 & 1.15 & 0.08 & 10.20 & 0.25 & 8.30 & 0.15 & 1.538 & 0.18 & -1.455 & 0.135 & 0.899 & 0.101\\
  UFMG66 & 260.852 & -32.050 & 200.0 & 21.7 & 0.80 & 0.05 & 10.50 & 0.20 & 7.55 & 0.15 & -1.107 & 0.128 & -2.465 & 0.156 & 0.591 & 0.040\\
  UFMG67 & 262.320 & -30.950 & 1041.7 & 62.9 & 0.38 & 0.05 & 10.15 & 0.15 & 8.30 & 0.20 & 3.806 & 0.158 & -0.296 & 0.107 & 0.955 & 0.060\\
  UFMG68 & 263.132 & -30.876 & 491.7 & 38.2 & 0.90 & 0.05 & 10.50 & 0.20 & 7.85 & 0.20 & -0.684 & 0.103 & -2.673 & 0.120 & 0.624 & 0.070\\
  UFMG69 & 264.500 & -29.500 & 666.7 & 38.2 & 1.90 & 0.12 & 11.00 & 0.35 & 8.20 & 0.20 & -0.426 & 0.287 & -0.220 & 0.111 & 0.284 & 0.103\\
  UFMG70 & 264.200 & -29.872 & 1041.7 & 62.9 & 0.75 & 0.08 & 10.60 & 0.35 & 8.90 & 0.15 & 1.446 & 0.126 & 0.568 & 0.204 & 0.529 & 0.060\\
  UFMG71 & 265.555 & -27.329 & 350.0 & 43.3 & 0.95 & 0.05 & 10.50 & 0.20 & 8.00 & 0.20 & 0.764 & 0.133 & -1.777 & 0.105 & 0.603 & 0.089\\
  UFMG72 & 268.515 & -23.170 & 708.3 & 38.2 & 0.75 & 0.10 & 10.10 & 0.30 & 8.10 & 0.20 & 0.096 & 0.137 & -2.569 & 0.124 & 0.686 & 0.049\\
  UFMG73 & 269.084 & -21.892 & 550.0 & 66.1 & 1.00 & 0.08 & 11.05 & 0.25 & 8.70 & 0.10 & 1.944 & 0.168 & 0.915 & 0.142 & 0.492 & 0.058\\
  UFMG74 & 269.736 & -20.588 & 350.0 & 43.3 & 1.35 & 0.10 & 11.20 & 0.40 & 8.75 & 0.10 & -0.435 & 0.158 & -1.785 & 0.143 & 0.289 & 0.089\\
  UFMG75 & 270.245 & -20.220 & 608.3 & 76.4 & 1.45 & 0.10 & 10.80 & 0.30 & 8.80 & 0.10 & 0.157 & 0.209 & -0.003 & 0.133 & 0.329 & 0.065\\
  UFMG76 & 261.175 & -34.690 & 666.7 & 38.2 & 1.90 & 0.10 & 10.70 & 0.30 & 6.90 & 0.25 & -1.181 & 0.216 & -3.022 & 0.171 & 0.571 & 0.083\\
  UFMG77 & 262.805 & -33.462 & 708.3 & 38.2 & 0.68 & 0.05 & 10.15 & 0.20 & 7.10 & 0.20 & 2.575 & 0.114 & -1.873 & 0.116 & 0.854 & 0.052\\
  UFMG78 & 268.613 & -24.891 & 550.0 & 66.1 & 0.80 & 0.10 & 11.00 & 0.25 & 8.90 & 0.15 & -0.623 & 0.132 & -3.008 & 0.192 & 0.377 & 0.057\\
  UFMG79 & 268.930 & -23.980 & 400.0 & 43.3 & 2.20 & 0.05 & 11.30 & 0.20 & 8.20 & 0.20 & -0.355 & 0.224 & -1.910 & 0.140 & 0.311 & 0.109\\
  UFMG80 & 270.188 & -22.188 & 1175.0 & 86.6 & 1.05 & 0.10 & 11.00 & 0.30 & 8.60 & 0.20 & 1.250 & 0.150 & 0.325 & 0.245 & 0.438 & 0.054\\
  UFMG81 & 270.156 & -22.431 & 400.0 & 43.3 & 1.40 & 0.15 & 10.85 & 0.35 & 8.60 & 0.20 & 0.359 & 0.170 & -1.760 & 0.131 & 0.410 & 0.070\\
  UFMG82 & 263.916 & -33.335 & 708.3 & 38.2 & 1.00 & 0.08 & 10.85 & 0.20 & 8.45 & 0.10 & -1.629 & 0.169 & -1.796 & 0.144 & 0.615 & 0.060\\
  UFMG83 & 268.085 & -28.174 & 758.3 & 62.9 & 0.78 & 0.12 & 11.15 & 0.40 & 8.80 & 0.25 & -0.101 & 0.247 & -2.035 & 0.264 & 0.300 & 0.053\\
  UFMG84 & 272.195 & -21.482 & 1175.0 & 86.6 & 0.85 & 0.07 & 10.70 & 0.30 & 7.25 & 0.30 & -0.625 & 0.165 & -1.420 & 0.134 & 0.604 & 0.066\\
  UFMG85 & 262.055 & -37.860 & 491.7 & 38.2 & 0.55 & 0.05 & 10.65 & 0.35 & 7.50 & 0.40 & 3.249 & 0.145 & -1.938 & 0.069 & 0.824 & 0.055\\
  UFMG86 & 261.578 & -37.800 & 1308.3 & 38.2 & 1.15 & 0.10 & 10.80 & 0.20 & 8.35 & 0.20 & 0.576 & 0.220 & -0.977 & 0.120 & 0.447 & 0.049\\
  UFMG87 & 263.283 & -35.135 & 1341.7 & 38.2 & 1.55 & 0.07 & 11.70 & 0.20 & 8.50 & 0.10 & -1.041 & 0.139 & -2.655 & 0.117 & 0.317 & 0.047\\
  UFMG88 & 268.210 & -29.431 & 816.7 & 62.9 & 0.58 & 0.05 & 10.50 & 0.25 & 8.70 & 0.10 & 1.960 & 0.464 & -1.563 & 0.259 & 0.588 & 0.086\\
  UFMG89 & 269.300 & -27.590 & 400.0 & 43.3 & 1.22 & 0.10 & 10.70 & 0.35 & 8.50 & 0.20 & 0.290 & 0.188 & -1.896 & 0.121 & 0.403 & 0.064\\
  UFMG90 & 270.655 & -25.830 & 608.3 & 76.4 & 1.15 & 0.07 & 10.85 & 0.35 & 8.90 & 0.10 & 0.455 & 0.227 & -0.083 & 0.244 & 0.440 & 0.025\\
  UFMG91 & 264.868 & -35.075 & 608.3 & 76.4 & 0.89 & 0.06 & 12.25 & 0.25 & 9.25 & 0.10 & 0.222 & 0.112 & -3.637 & 0.113 & 0.269 & 0.031\\
  UFMG92 & 269.039 & -29.494 & 491.7 & 38.2 & 0.70 & 0.08 & 10.95 & 0.20 & 8.35 & 0.20 & -0.094 & 0.146 & -0.458 & 0.118 & 0.390 & 0.073\\
  UFMG93 & 269.464 & -29.155 & 400.0 & 43.3 & 0.65 & 0.10 & 11.40 & 0.35 & 9.00 & 0.10 & -0.091 & 0.151 & -2.509 & 0.112 & 0.335 & 0.058\\
  UFMG94 & 267.959 & -33.390 & 708.3 & 38.2 & 0.70 & 0.05 & 10.30 & 0.20 & 8.35 & 0.15 & 0.495 & 0.087 & -2.807 & 0.086 & 0.782 & 0.049\\
  UFMG95 & 270.695 & -29.370 & 708.3 & 38.2 & 0.15 & 0.07 & 10.00 & 0.25 & 8.20 & 0.20 & 1.815 & 0.153 & -0.895 & 0.124 & 0.839 & 0.056\\
  UFMG96 & 264.455 & -29.776 & 866.7 & 38.2 & 1.28 & 0.07 & 10.90 & 0.25 & 8.55 & 0.10 & -1.442 & 0.140 & -3.421 & 0.151 & 0.413 & 0.054\\
\hline
\end{tabular}
\end{table*}

\section{Analysis}
\label{sect:analysis}

\subsection{Radial density profiles}
\label{sect:4.2}

We used the peak values of proper motions and parallaxes computed in the Sect. \ref{sect:search} and restricted our sample by a 3D box in the astrometric space: 1\,mas\,yr$^{-1}$ around the mean proper motion values and 1 mas around the mean parallax. 

Then we used the centre coordinates as a first guess to build the Radial Density Profiles (RDPs). The RDPs were used to estimate the size of our OC candidates, the local background density and to refine the centres previously computed. See F+2020 for details.
We defined the limiting radius ($r_{lim}$) as the radius beyond which stellar densities start do fluctuate around a nearly constant value. 

\begin{figure*}
\centering
\includegraphics[width=0.24\linewidth]{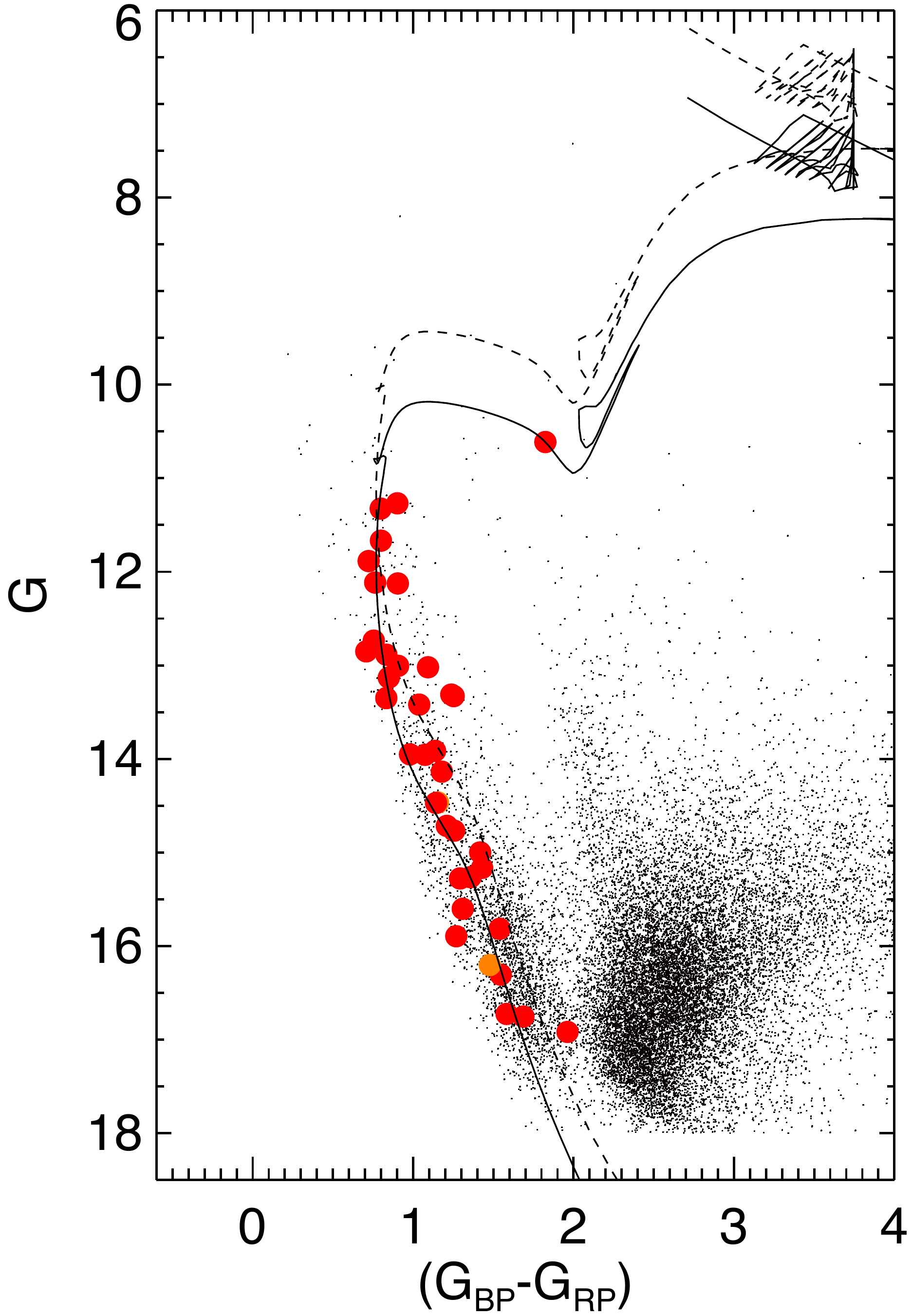} \hspace{0.0cm}
\includegraphics[width=0.27\linewidth]{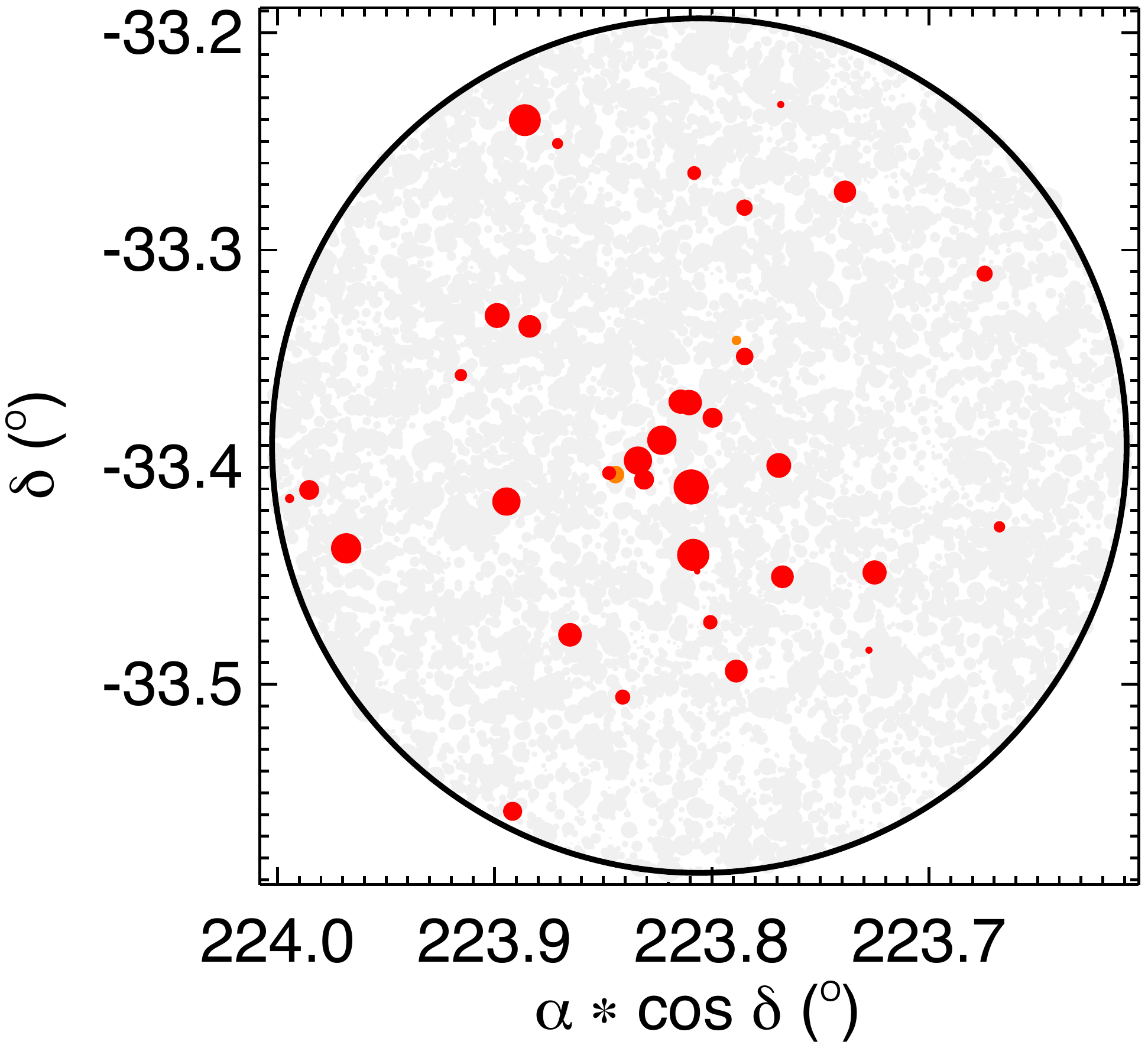}  \hspace{0.00cm}
\includegraphics[width=0.36\linewidth]{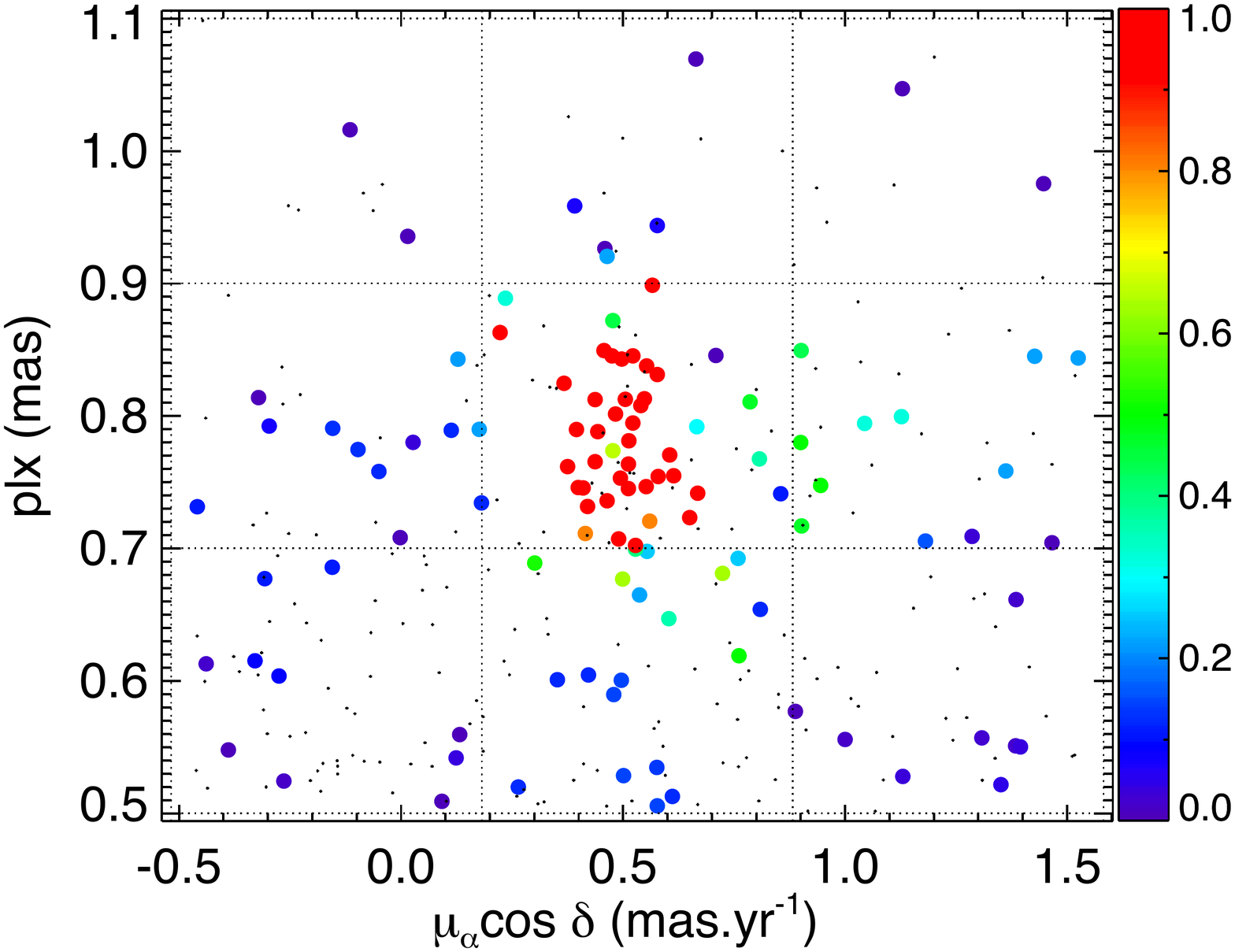} \hspace{0.0cm}


\caption{Sequence of panels showing our results for the cluster UFMG94. Left: example of PARSEC-COLIBRI isochrone fitted (solid line) over the cleaned CMD. We overplotted the corresponding binary sequence (dashed lines) by deducing 0.75 mag from the G magnitude values. Middle: and example of the cluster members spatial distribution. Right: $\varpi$ versus $\mu_{\alpha}\,\textrm{cos}\,\delta$ plot for stars in the cluster area (that is, $r\,\leq\,r_{\textrm{lim}}$). In each panel, the symbol colours were assigned according to the membership scale represented by the colour bar in the rightmost plot.}
\label{fig:isoc1}
\end{figure*}

\subsection{Assessing membership and isochrone fittings}
\label{sect:sec4}
Since our cluster candidates are projected against dense stellar fields, in order to derive stars membership likelihoods, we applied a routine (fully described in \citeauthor{2019A&A...624A...8A}\,\,\citeyear{2019A&A...624A...8A}) that evaluates statistically the overdensity of cluster 
stars in comparison to those in a nearby field in the 3D astrometric space ($\mu_{\alpha}^{*}$, $\mu_{\delta}$, $\varpi$).   
We call the region centred in the cluster's coordinates and restricted by its $r_{\mathrm{lim}}$ as the cluster region. The comparison field is restricted by a ring-like region with inner radius  $r_{\mathrm{lim}}$+3\arcmin. The outer radius is such that the field area is equal to 3 times the cluster region. A detailed description of the procedure can be find in F+2020.
 We employed PARSEC-COLIBRI models \citep{Marigo:2017} 
to perform isochrone fittings on the CMDs of the decontaminated cluster samples to determine age, distance and reddening. To do so, we visually inspected the match between the model and the cluster stars loci in specific evolutionary regions (main sequence, the turnoff, and the giant clump). To convert $E(G_{BP}-G_{RP})$ to $E(B-V)$, we adopted a reddening law \citep{Cardelli:1989,Odonnell:1994}.
Fig.~\ref{fig:isoc1} presents a sequence of panels showing our results for the cluster UFMG94: the best-fitting isochrone to the decontaminated CMD, the cluster members spatial distribution and a diagram showing the cluster members distribution in proper motion in declination in function of their parallaxes.

The resulting values for the centre coordinate, size, mean astrometric values, colour excess, distance module and age for all the clusters are presented in the Table \ref{Tab:clusters_prop1}. This table is also available electronically. 
 We have crossmatched our memberlist catalogue ids with \textit{Gaia} EDR3 \citep{2020arXiv201201533G}, applied the quality filters that ensures astrometric valid solutions and made the zero point parallaxes correction. The OCs 3d astrometric spaces ($\mu_{\alpha}^{*}$, $\mu_{\delta}$ and $\varpi$) are concentrated and present similar mean proper motions, paralaxes and dispersions with respect to \textit{Gaia} DR2, i.e., the OC candidates presented a concentrated structure and seems not to be random fluctuations in \textit{Gaia} DR2 data.

\section{Discussion}
\label{sect:sec5}


According to the literature (\citealp{Dias:2002,Kharchenko:2013,10.1093/mnras/staa1684,sla19,lp19,cjl20,ca20,2020arXiv201014870H}), there exist at least 206 OCs with determined parameters projected towards the Galactic centre direction. Our method was capable to recover 109 OCs in this region, 34 of them are reported as new objects, which represents an increment of $\sim 17\%$ of the known OCs with derived parameters in this same direction. 



According to the mentioned catalogues, there exist 82 OCs with known distance values within 2\,kpc. Our new OCs sample represents a fractional increment of $\sim 19\%$ up to 1.1\,kpc, 38\% up to 1.5\,kpc and 39\% up to 2\,kpc. According to the catalogues, 118 OCs have astrometric parameters from \textit{Gaia} DR2 data in this region, in this way, our findings represent an increment of about $\sim 30\%$ of this sample. The increment represented by our newly found OCs is shown in the Table\ \ref{tab:estatistica}.



\begin{figure}
\centering
\includegraphics[width=0.62\linewidth]{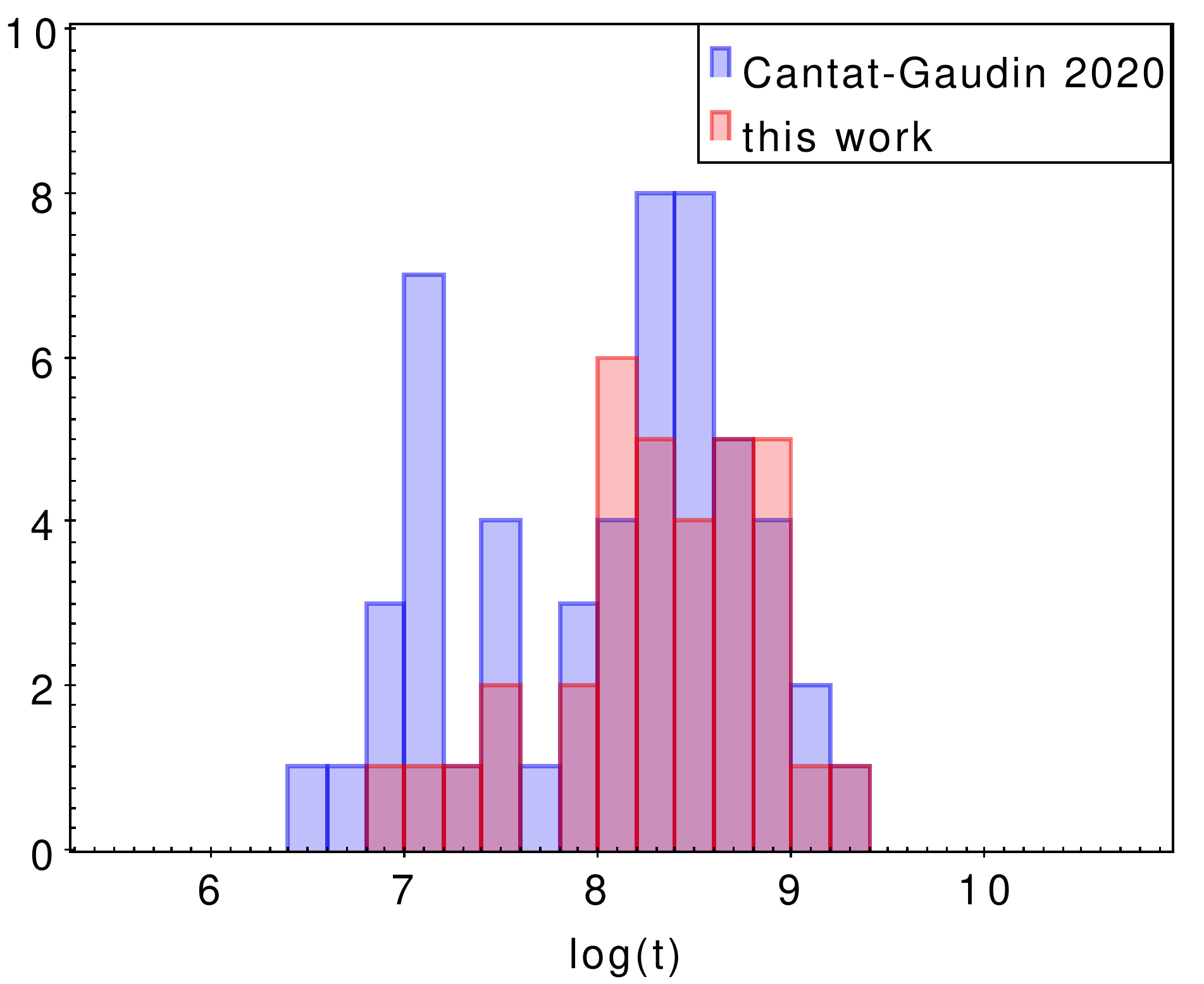} \hspace{0.25cm}
\caption{Blue: OC sample from the homogeneous OC catalogue constituted by objects characterized with \textit{Gaia} DR2 data presented in CG+2020 and restricted by the Galactic coordinates $-10^{\circ}$ $\leq \ell \leq$10$^{\circ}$ and $-5^{\circ}$ $\leq b \leq$5$^{\circ}$. Red: The OC sample found in this work.}
\label{fig:cantat_ferreira_histo}
\end{figure}

\begin{figure}
\centering
\includegraphics[width=0.97\linewidth]{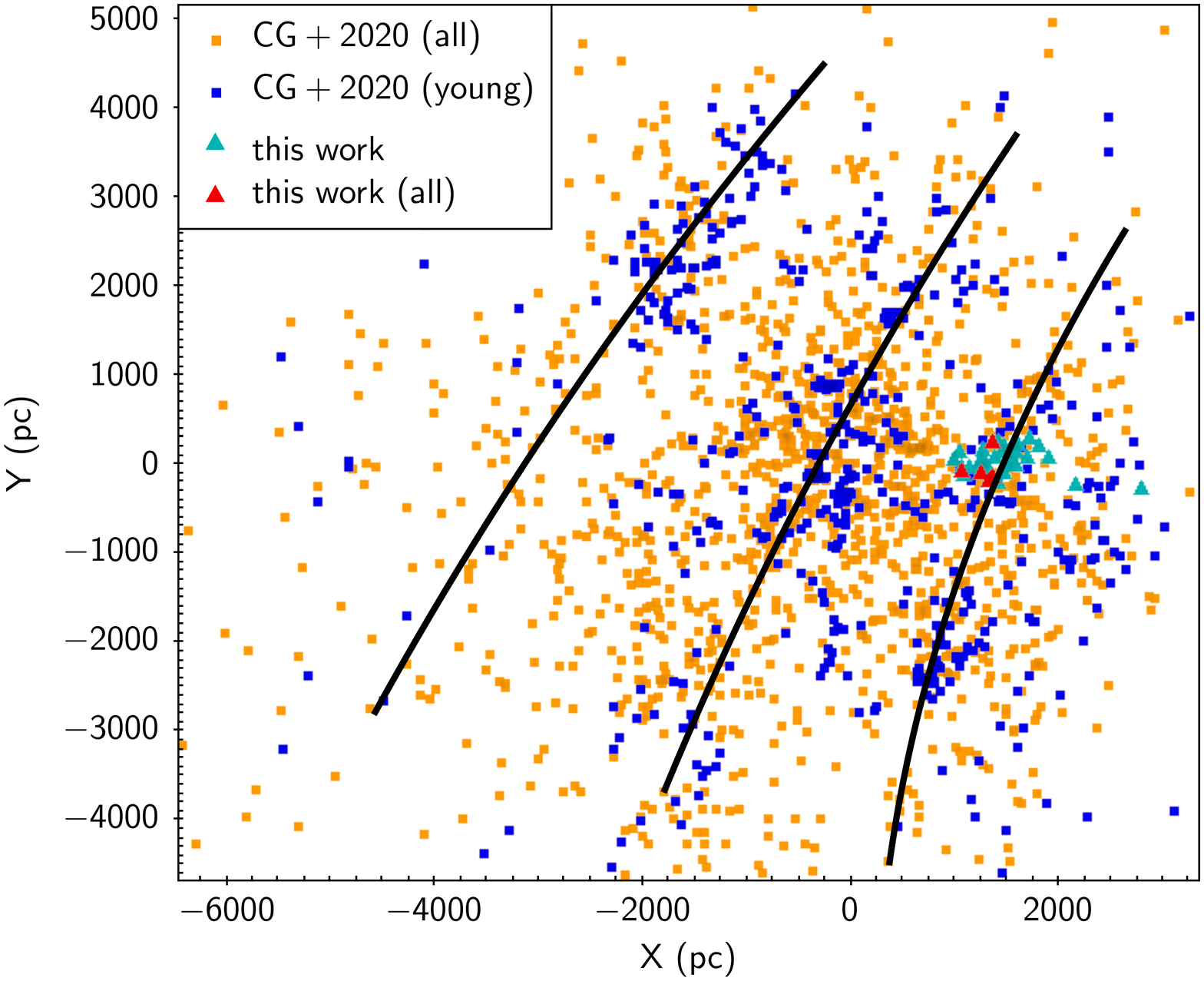} \hspace{0.2cm}
\includegraphics[width=0.87\linewidth]{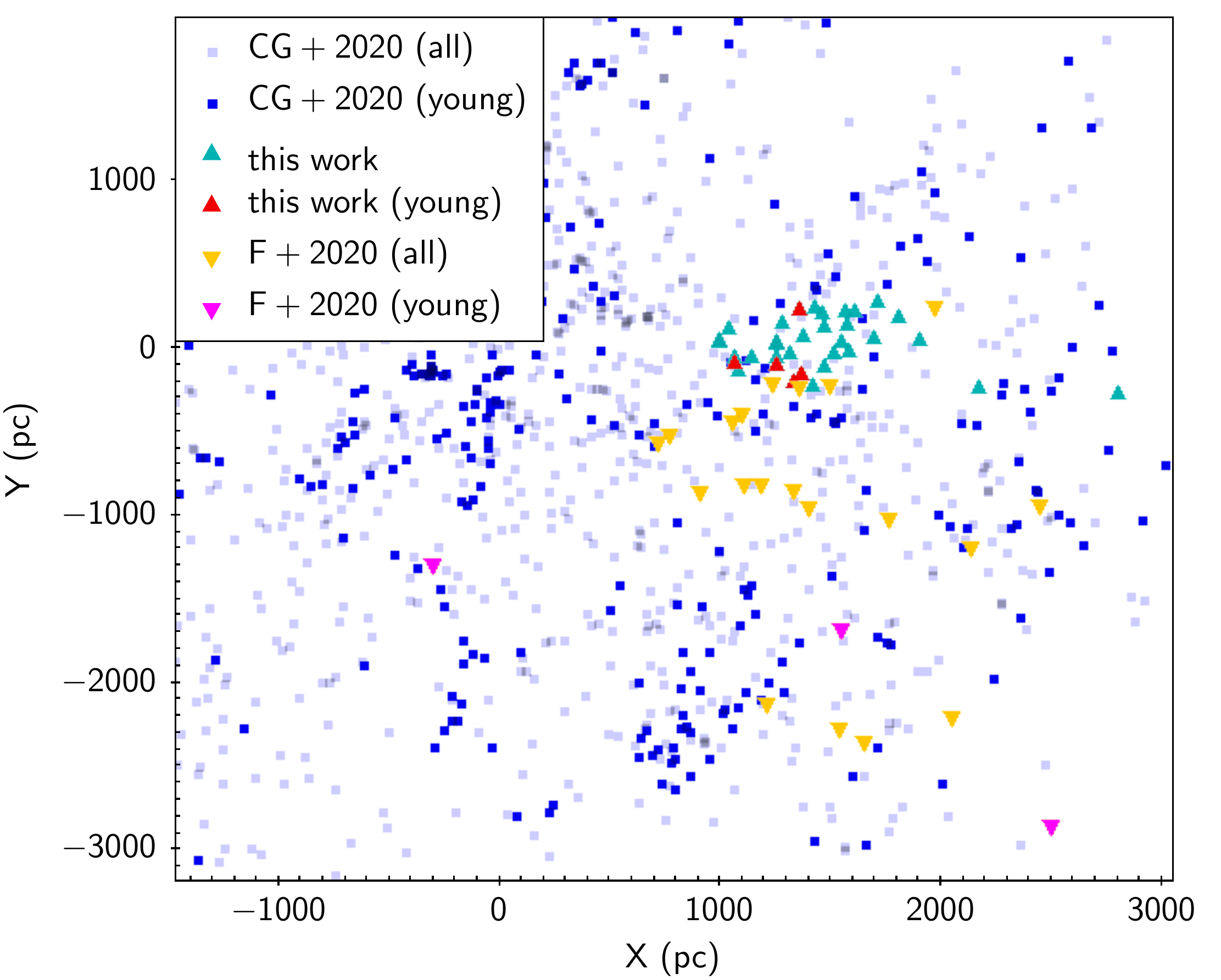} \hspace{0.3cm}
\caption{Top: OCs presented in CG+2020 (orange and red squares) and in the present work (cyan and blue triangles). Bottom: zoom in showing also objects from F+2020 (yellow and pink upside down triangles).}
\label{fig:cantat_ferreira_arm}
\end{figure}

Recently, CG+2020 built a large and homogeneous OC catalogue constituted by objects characterized with \textit{Gaia} DR2 data, which includes known OCs and newly found OCs. Their catalogue contains 60 OCs within the region covered in this work, which means that our sample provides a significant increase of objects with distance, colour excess and age in the investigated region. Fig. \ref{fig:cantat_ferreira_arm} shows OCs from CG+2020 (orange and red square) and our newly discovered sample (cyan and red triangles) in the Galactic plane.
Fig. \ref{fig:cantat_ferreira_histo} shows that most of our new findings present ages  between $8.0 < \textrm{log}\,t < 9.0$.



The top panel of the Fig. \ref{fig:cantat_ferreira_arm} shows how young OCs (blue squares), with ages limited by 50\,Myr (corresponding to $\log{t}<7.7$), delineate the local Galactic structure as spiral arms. We note that our findings are located in the Sa\-gi\-tta\-rius arm, with the exception of two objects (UFMG 87 and UFMG91).



The bottom panel of the Fig. \ref{fig:cantat_ferreira_arm} shows how our findings  contributed to the local OCs population in the Sagittarius arm. However, even with the addition of objects presented in F+2020, we still note a lack of objects at the position $(X,Y)\sim (1000,-1000)$\,pc, suggesting that either the arm is discontinuous at this position or we may still be missing young clusters at this particular direction.

\begin{table}
\centering
\footnotesize
\caption{The impact of the newly discovered OCs in this work over the local OC population.}
\def\arraystretch{1.3}
\begin{tabular}{l c c c} \hline
Sample & fraction & increase &  \\
\hline
Total & $34/206$ & $17\%$ &  \\
with \textit{Gaia} data & $34/118$ & $29\%$ &  \\
with astrophysical parameters & $34/129$ & $26\%$ &  \\
with $d<2\ kpc$ & $32/82$  & $39\%$ &  \\
with $d<1.5\ kpc$ & $22/57$ & $38\%$ &  \\
with $d<1.1\ kpc$ & $6/32$ & $19\%$ &  \\

\hline
\end{tabular}
\def\arraystretch{1.0}
\label{tab:estatistica}
\end{table}



\section{Summary and concluding remarks}
\label{sect:sec7}

We discovered and derived astrophysical parameters for 34 OCs projected in the Galactic bulge direction using \textit{Gaia} DR2 data. Our methodology, combined with the high precision of \textit{Gaia} DR2 astrometric and photometric data, allowed us to find the OCs in the astrometric space. The investigated OCs in this work are mainly located within 2 kpc from the Sun and are projected in the Galactic bulge direction. Their ages are comprised between $\sim$10 Myr and $\sim$1 Gyr and colour excesses between $0.1$ and $2.3$.

When restricting the distance range to 2\,kpc, our OCs sample represents an increment of $\sim 39\%$ at this regime of distances. In the same way, our findings represent an increment of about $\sim 30\%$ of OCs that have been characterized by means of \textit{Gaia} DR2 data.
Although these objects fill an important gap in the open clusters distribution along the Sagittarius arm, this arm, traced by known
clusters, appears to be interrupted. Even though such discontinuities have been observed in other galaxies, it could also be an artifact due to the incompleteness of the cluster census.

\section*{Acknowledgements}

We thank the agencies FAPEMIG, CNPq and CAPES (finance code 001). This research has made use of the VizieR catalogue access tool, CDS, Strasbourg, France and has made use of data from the European Space Agency (ESA) mission \textit{Gaia} (\url{https://www.cosmos.esa.int/gaia}), processed by the \textit{Gaia} Data Processing and Analysis Consortium (DPAC, \url{https://www.cosmos.esa.int/web/gaia/dpac/consortium}). Funding for the DPAC has been provided by national institutions, in particular the institutions participating in the \textit{Gaia} Multilateral Agreement. This  research  has  made  use  of  TOPCAT \citep{Taylor:2005}.

\section*{Data availability}
The data underlying this article is publicly available (Gaia DR2) or is available in the article.


\bibliographystyle{mnras}
\bibliography{references} 

\begin{thebibliography}{}
\makeatletter
\relax
\def\mn@urlcharsother{\let\do\@makeother \do\$\do\&\do\#\do\^\do\_\do\%\do\~}
\def\mn@doi{\begingroup\mn@urlcharsother \@ifnextchar [ {\mn@doi@}
  {\mn@doi@[]}}
\def\mn@doi@[#1]#2{\def\@tempa{#1}\ifx\@tempa\@empty \href
  {http://dx.doi.org/#2} {doi:#2}\else \href {http://dx.doi.org/#2} {#1}\fi
  \endgroup}
\def\mn@eprint#1#2{\mn@eprint@#1:#2::\@nil}
\def\mn@eprint@arXiv#1{\href {http://arxiv.org/abs/#1} {{\tt arXiv:#1}}}
\def\mn@eprint@dblp#1{\href {http://dblp.uni-trier.de/rec/bibtex/#1.xml}
  {dblp:#1}}
\def\mn@eprint@#1:#2:#3:#4\@nil{\def\@tempa {#1}\def\@tempb {#2}\def\@tempc
  {#3}\ifx \@tempc \@empty \let \@tempc \@tempb \let \@tempb \@tempa \fi \ifx
  \@tempb \@empty \def\@tempb {arXiv}\fi \@ifundefined
  {mn@eprint@\@tempb}{\@tempb:\@tempc}{\expandafter \expandafter \csname
  mn@eprint@\@tempb\endcsname \expandafter{\@tempc}}}

\bibitem[\protect\citeauthoryear{{Angelo}, {Santos}, {Corradi}  \&
  {Maia}}{{Angelo} et~al.}{2019}]{2019A&A...624A...8A}
{Angelo} M.~S.,  {Santos} J.~F.~C.,  {Corradi} W.~J.~B.,   {Maia} F.~F.~S.,
  2019, \mn@doi [\aap] {10.1051/0004-6361/201832702}, \href
  {https://ui.adsabs.harvard.edu/abs/2019A&A...624A...8A} {624, A8}

\bibitem[\protect\citeauthoryear{Angelo, Corradi, Santos, Maia  \&
  Ferreira}{Angelo et~al.}{2020}]{10.1093/mnras/staa3192}
Angelo M.~S.,  Corradi W. J.~B.,  Santos J~F~C J.,  Maia F. F.~S.,   Ferreira
  F.~A.,  2020, \mn@doi [\mnras] {10.1093/mnras/staa3192}, 500, 4338

\bibitem[\protect\citeauthoryear{{Arenou} et~al.,}{{Arenou}
  et~al.}{2018}]{Arenou:2018}
{Arenou} F.,  et~al., 2018, \mn@doi [\aap] {10.1051/0004-6361/201833234}, \href
  {http://adsabs.harvard.edu/abs/2018A%26A...616A..17A} {616, A17}

\bibitem[\protect\citeauthoryear{{Bastian, U.}}{{Bastian, U.}}{2019}]{refId0}
{Bastian, U.} 2019, \mn@doi [A\&A] {10.1051/0004-6361/201936595}, 630, L8

\bibitem[\protect\citeauthoryear{{Bica}, {Pavani}, {Bonatto}  \& {Lima}}{{Bica}
  et~al.}{2019}]{2019AJ....157...12B}
{Bica} E.,  {Pavani} D.~B.,  {Bonatto} C.~J.,   {Lima} E.~F.,  2019, \mn@doi
  [\aj] {10.3847/1538-3881/aaef8d}, \href
  {https://ui.adsabs.harvard.edu/abs/2019AJ....157...12B} {157, 12}

\bibitem[\protect\citeauthoryear{{Borissova} et~al.,}{{Borissova}
  et~al.}{2018}]{2018MNRAS.481.3902B}
{Borissova} J.,  et~al., 2018, \mn@doi [\mnras] {10.1093/mnras/sty2354}, \href
  {https://ui.adsabs.harvard.edu/abs/2018MNRAS.481.3902B} {481, 3902}

\bibitem[\protect\citeauthoryear{{Cantat-Gaudin} \& {Anders}}{{Cantat-Gaudin}
  \& {Anders}}{2020}]{ca20}
{Cantat-Gaudin} T.,  {Anders} F.,  2020, \mn@doi [\aap]
  {10.1051/0004-6361/201936691}, \href
  {https://ui.adsabs.harvard.edu/abs/2020A&A...633A..99C} {633, A99}

\bibitem[\protect\citeauthoryear{{Cantat-Gaudin} et~al.,}{{Cantat-Gaudin}
  et~al.}{2018}]{cjv18}
{Cantat-Gaudin} T.,  et~al., 2018, \mn@doi [\aap]
  {10.1051/0004-6361/201833476}, \href
  {https://ui.adsabs.harvard.edu/abs/2018A&A...618A..93C} {618, A93}

\bibitem[\protect\citeauthoryear{{Cantat-Gaudin} et~al.,}{{Cantat-Gaudin}
  et~al.}{2019}]{cks19}
{Cantat-Gaudin} T.,  et~al., 2019, \mn@doi [\aap]
  {10.1051/0004-6361/201834453}, \href
  {https://ui.adsabs.harvard.edu/abs/2019A&A...624A.126C} {624, A126}

\bibitem[\protect\citeauthoryear{{Cantat-Gaudin} et~al.,}{{Cantat-Gaudin}
  et~al.}{2020}]{2020A&A...640A...1C}
{Cantat-Gaudin} T.,  et~al., 2020, \mn@doi [\aap]
  {10.1051/0004-6361/202038192}, \href
  {https://ui.adsabs.harvard.edu/abs/2020A&A...640A...1C} {640, A1}

\bibitem[\protect\citeauthoryear{{Cardelli}, {Clayton}  \& {Mathis}}{{Cardelli}
  et~al.}{1989}]{Cardelli:1989}
{Cardelli} J.~A.,  {Clayton} G.~C.,   {Mathis} J.~S.,  1989, \mn@doi [\apj]
  {10.1086/167900}, \href {http://adsabs.harvard.edu/abs/1989ApJ...345..245C}
  {345, 245}

\bibitem[\protect\citeauthoryear{{Castro-Ginard}, {Jordi}, {Luri}, {Julbe},
  {Morvan}, {Balaguer-N{\'u}{\~n}ez}  \& {Cantat-Gaudin}}{{Castro-Ginard}
  et~al.}{2018}]{cjl18}
{Castro-Ginard} A.,  {Jordi} C.,  {Luri} X.,  {Julbe} F.,  {Morvan} M.,
  {Balaguer-N{\'u}{\~n}ez} L.,   {Cantat-Gaudin} T.,  2018, \mn@doi [\aap]
  {10.1051/0004-6361/201833390}, \href
  {https://ui.adsabs.harvard.edu/abs/2018A&A...618A..59C} {618, A59}

\bibitem[\protect\citeauthoryear{{Castro-Ginard}, {Jordi}, {Luri},
  {Cantat-Gaudin}  \& {Balaguer-N{\'u}{\~n}ez}}{{Castro-Ginard}
  et~al.}{2019}]{cjl19}
{Castro-Ginard} A.,  {Jordi} C.,  {Luri} X.,  {Cantat-Gaudin} T.,
  {Balaguer-N{\'u}{\~n}ez} L.,  2019, \mn@doi [\aap]
  {10.1051/0004-6361/201935531}, \href
  {https://ui.adsabs.harvard.edu/abs/2019A&A...627A..35C} {627, A35}

\bibitem[\protect\citeauthoryear{{Castro-Ginard} et~al.,}{{Castro-Ginard}
  et~al.}{2020}]{cjl20}
{Castro-Ginard} A.,  et~al., 2020, \mn@doi [\aap]
  {10.1051/0004-6361/201937386}, \href
  {https://ui.adsabs.harvard.edu/abs/2020A&A...635A..45C} {635, A45}

\bibitem[\protect\citeauthoryear{{Dias}, {Alessi}, {Moitinho}  \&
  {L{\'e}pine}}{{Dias} et~al.}{2002}]{Dias:2002}
{Dias} W.~S.,  {Alessi} B.~S.,  {Moitinho} A.,   {L{\'e}pine} J.~R.~D.,  2002,
  \mn@doi [\aap] {10.1051/0004-6361:20020668}, \href
  {http://adsabs.harvard.edu/abs/2002A%26A...389..871D} {389, 871}

\bibitem[\protect\citeauthoryear{{Dias}, {Monteiro}, {L{\'e}pine}  \&
  {Barros}}{{Dias} et~al.}{2019}]{2019MNRAS.486.5726D}
{Dias} W.~S.,  {Monteiro} H.,  {L{\'e}pine} J.~R.~D.,   {Barros} D.~A.,  2019,
  \mn@doi [\mnras] {10.1093/mnras/stz1196}, \href
  {https://ui.adsabs.harvard.edu/abs/2019MNRAS.486.5726D} {486, 5726}

\bibitem[\protect\citeauthoryear{{Evans} et~al.,}{{Evans}
  et~al.}{2018}]{Evans:2018}
{Evans} D.~W.,  et~al., 2018, \mn@doi [\aap] {10.1051/0004-6361/201832756},
  \href {https://ui.adsabs.harvard.edu/#abs/2018A&A...616A...4E} {616, A4}

\bibitem[\protect\citeauthoryear{{Ferreira}, {Santos}, {Corradi}, {Maia}  \&
  {Angelo}}{{Ferreira} et~al.}{2019}]{2019MNRAS.483.5508F}
{Ferreira} F.~A.,  {Santos} J.~F.~C.,  {Corradi} W.~J.~B.,  {Maia} F.~F.~S.,
  {Angelo} M.~S.,  2019, \mn@doi [\mnras] {10.1093/mnras/sty3511}, \href
  {http://adsabs.harvard.edu/abs/2019MNRAS.483.5508F} {483, 5508}

\bibitem[\protect\citeauthoryear{Ferreira, Corradi, Maia, Angelo  \&
  Santos}{Ferreira et~al.}{2020}]{10.1093/mnras/staa1684}
Ferreira F.~A.,  Corradi W. J.~B.,  Maia F. F.~S.,  Angelo M.~S.,   Santos
  J~F~C J.,  2020, \mn@doi [\mnras] {10.1093/mnras/staa1684}, 496, 2021

\bibitem[\protect\citeauthoryear{{Gaia Collaboration} et~al.,}{{Gaia
  Collaboration} et~al.}{2018}]{Brown:2018}
{Gaia Collaboration} et~al., 2018, \mn@doi [\aap]
  {10.1051/0004-6361/201833051}, \href
  {http://adsabs.harvard.edu/abs/2018A%26A...616A...1G} {616, A1}

\bibitem[\protect\citeauthoryear{{Gaia Collaboration}, {Brown}, {Vallenari},
  {Prusti}, {de Bruijne}, {Babusiaux}  \& {Biermann}}{{Gaia Collaboration}
  et~al.}{2020}]{2020arXiv201201533G}
{Gaia Collaboration} {Brown} A.~G.~A.,  {Vallenari} A.,  {Prusti} T.,  {de
  Bruijne} J.~H.~J.,  {Babusiaux} C.,   {Biermann} M.,  2020, arXiv e-prints,
  \href {https://ui.adsabs.harvard.edu/abs/2020arXiv201201533G} {p.
  arXiv:2012.01533}

\bibitem[\protect\citeauthoryear{{Guo} et~al.,}{{Guo}
  et~al.}{2018}]{2018RAA....18...32G}
{Guo} J.-C.,  et~al., 2018, \mn@doi [RAA] {10.1088/1674-4527/18/3/32}, \href
  {https://ui.adsabs.harvard.edu/abs/2018RAA....18...32G} {18, 032}

\bibitem[\protect\citeauthoryear{{Hao}, {Xu}, {Wu}, {He}  \& {Bian}}{{Hao}
  et~al.}{2020}]{2020PASP..132c4502H}
{Hao} C.,  {Xu} Y.,  {Wu} Z.,  {He} Z.,   {Bian} S.,  2020, \mn@doi [\pasp]
  {10.1088/1538-3873/ab694d}, \href
  {https://ui.adsabs.harvard.edu/abs/2020PASP..132c4502H} {132, 034502}

\bibitem[\protect\citeauthoryear{{He}, {Xu}, {Hao}, {Wu}  \& {Li}}{{He}
  et~al.}{2020}]{2020arXiv201014870H}
{He} Z.-H.,  {Xu} Y.,  {Hao} C.-J.,  {Wu} Z.-Y.,   {Li} J.-J.,  2020, arXiv
  e-prints, \href {https://ui.adsabs.harvard.edu/abs/2020arXiv201014870H} {p.
  arXiv:2010.14870}

\bibitem[\protect\citeauthoryear{{Kharchenko}, {Piskunov}, {Schilbach},
  {R{\"o}ser}  \& {Scholz}}{{Kharchenko} et~al.}{2013}]{Kharchenko:2013}
{Kharchenko} N.~V.,  {Piskunov} A.~E.,  {Schilbach} E.,  {R{\"o}ser} S.,
  {Scholz} R.-D.,  2013, \mn@doi [\aap] {10.1051/0004-6361/201322302}, \href
  {http://adsabs.harvard.edu/abs/2013A%26A...558A..53K} {558, A53}

\bibitem[\protect\citeauthoryear{{Lindegren} et~al.,}{{Lindegren}
  et~al.}{2018}]{Lindegren:2018}
{Lindegren} L.,  et~al., 2018, \mn@doi [\aap] {10.1051/0004-6361/201832727},
  \href {http://adsabs.harvard.edu/abs/2018A%26A...616A...2L} {616, A2}

\bibitem[\protect\citeauthoryear{{Liu} \& {Pang}}{{Liu} \& {Pang}}{2019}]{lp19}
{Liu} L.,  {Pang} X.,  2019, \mn@doi [\apjs] {10.3847/1538-4365/ab530a}, \href
  {https://ui.adsabs.harvard.edu/abs/2019ApJS..245...32L} {245, 32}

\bibitem[\protect\citeauthoryear{{Magrini} et~al.,}{{Magrini}
  et~al.}{2017}]{2017A&A...603A...2M}
{Magrini} L.,  et~al., 2017, \mn@doi [\aap] {10.1051/0004-6361/201630294},
  \href {https://ui.adsabs.harvard.edu/abs/2017A&A...603A...2M} {603, A2}

\bibitem[\protect\citeauthoryear{{Marigo} et~al.,}{{Marigo}
  et~al.}{2017}]{Marigo:2017}
{Marigo} P.,  et~al., 2017, \mn@doi [\apj] {10.3847/1538-4357/835/1/77}, \href
  {http://adsabs.harvard.edu/abs/2017ApJ...835...77M} {835, 77}

\bibitem[\protect\citeauthoryear{{Monteiro}, {Dias}, {Moitinho},
  {Cantat-Gaudin}, {L{\'e}pine}, {Carraro}  \& {Paunzen}}{{Monteiro}
  et~al.}{2020}]{2020MNRAS.499.1874M}
{Monteiro} H.,  {Dias} W.~S.,  {Moitinho} A.,  {Cantat-Gaudin} T.,
  {L{\'e}pine} J.~R.~D.,  {Carraro} G.,   {Paunzen} E.,  2020, \mn@doi [\mnras]
  {10.1093/mnras/staa2983}, \href
  {https://ui.adsabs.harvard.edu/abs/2020MNRAS.499.1874M} {499, 1874}

\bibitem[\protect\citeauthoryear{{O'Donnell}}{{O'Donnell}}{1994}]{Odonnell:1994}
{O'Donnell} J.~E.,  1994, \mn@doi [\apj] {10.1086/173713}, \href
  {http://adsabs.harvard.edu/abs/1994ApJ...422..158O} {422, 158}

\bibitem[\protect\citeauthoryear{{Qin}, {Li}, {Chen}  \& {Zhong}}{{Qin}
  et~al.}{2020}]{2020arXiv200807164Q}
{Qin} S.-m.,  {Li} J.,  {Chen} L.,   {Zhong} J.,  2020, arXiv e-prints, \href
  {https://ui.adsabs.harvard.edu/abs/2020arXiv200807164Q} {p. arXiv:2008.07164}

\bibitem[\protect\citeauthoryear{{Rangwal}, {Yadav}, {Durgapal}, {Bisht}  \&
  {Nardiello}}{{Rangwal} et~al.}{2019}]{2019MNRAS.490.1383R}
{Rangwal} G.,  {Yadav} R.~K.~S.,  {Durgapal} A.,  {Bisht} D.,   {Nardiello} D.,
   2019, \mn@doi [\mnras] {10.1093/mnras/stz2642}, \href
  {https://ui.adsabs.harvard.edu/abs/2019MNRAS.490.1383R} {490, 1383}

\bibitem[\protect\citeauthoryear{{R{\"o}ser}, {Schilbach}  \&
  {Goldman}}{{R{\"o}ser} et~al.}{2016}]{Roser:2016}
{R{\"o}ser} S.,  {Schilbach} E.,   {Goldman} B.,  2016, \mn@doi [\aap]
  {10.1051/0004-6361/201629158}, \href
  {http://adsabs.harvard.edu/abs/2016A%26A...595A..22R} {595, A22}

\bibitem[\protect\citeauthoryear{{Sim}, {Lee}, {Ann}  \& {Kim}}{{Sim}
  et~al.}{2019}]{sla19}
{Sim} G.,  {Lee} S.~H.,  {Ann} H.~B.,   {Kim} S.,  2019, \mn@doi [JKAS]
  {10.5303/JKAS.2019.52.5.145}, \href
  {https://ui.adsabs.harvard.edu/abs/2019JKAS...52..145S} {52, 145}

\bibitem[\protect\citeauthoryear{{Taylor}}{{Taylor}}{2005}]{Taylor:2005}
{Taylor} M.~B.,  2005, in {Shopbell} P.,  {Britton} M.,   {Ebert} R.,  eds,
  ~ASPC Vol. 347, Astronomical Data Analysis Software and Systems XIV. p.~29

\bibitem[\protect\citeauthoryear{{Torrealba}, {Belokurov}  \&
  {Koposov}}{{Torrealba} et~al.}{2019}]{2019MNRAS.484.2181T}
{Torrealba} G.,  {Belokurov} V.,   {Koposov} S.~E.,  2019, \mn@doi [\mnras]
  {10.1093/mnras/stz071}, \href
  {https://ui.adsabs.harvard.edu/abs/2019MNRAS.484.2181T} {484, 2181}

\makeatother
\end{thebibliography}

\bsp	
\label{lastpage}
\end{document}